\documentclass[fleqn,twoside]{article}
\usepackage{espcrc2}

\usepackage{graphicx}
\usepackage[figuresright]{rotating}

\newcommand{\AmS}{{\protect\the\textfont2
  A\kern-.1667em\lower.5ex\hbox{M}\kern-.125emS}}

\title{Search for photons at the Pierre Auger Observatory}

\author{M.~Risse\address[BUW]{Bergische Universit\"at Wuppertal, 
        42119 Wuppertal, Germany}\thanks{Currently at University of Siegen,
              Germany} for the Pierre Auger Collaboration
}

\begin{document}

\begin{abstract}
The Pierre Auger Observatory has a unique potential to search for ultra-high energy
photons (above $\sim$1~EeV). First experimental limits on photons were obtained
during construction of the southern part of the Observatory.
Remarkably, already these limits have proven useful to falsify proposals about the
origin of cosmic rays, and to perform fundamental physics by constraining Lorentz
violation. A final discovery of photons at the upper end of the electromagnetic spectrum
is likely to impact various branches of physics and astronomy.
\vspace{1pc}
\end{abstract}

\maketitle

\section{Introduction}

There are good reasons to search for ultra-high energy
(UHE, above $\sim$10$^{18}$~eV = 1~EeV)
photons at the Auger Observatory~\cite{auger}.
An UHE photon produces a ``normal''
air shower, easily detectable with a giant shower observatory. Still, 
such photon showers have characteristics that make them well distinguishable
from showers initiated by primary hadrons. The calculation of these
photon shower characteristics can be done with high confidence because QED
effects dominate, such that solid conclusions can be drawn from the data.
A substantial flux of UHE photons is predicted in top-down models~\cite{topdown}.
Even in conventional cosmic-ray scenarios, UHE photons may be produced
at detectable level~\cite{gzk-photons};
as their expected flux at Earth depends on uncertain parameters such
as source density, injection spectrum or extragalactic radio background,
findings on UHE photons will provide valuable astrophysics information.

In the following, the current status of the search for UHE photons at the
Pierre Auger Observatory is briefly summarized. Detailed descriptions of the
data analyses can be found in Refs.~\cite{augerfd,augersd}. A general review
on the search for UHE photons using air showers is given in Ref.~\cite{review}.

\section{Search with hybrid data}

\begin{figure}[tb]
\vspace{9pt}
\includegraphics[width=17.5pc]{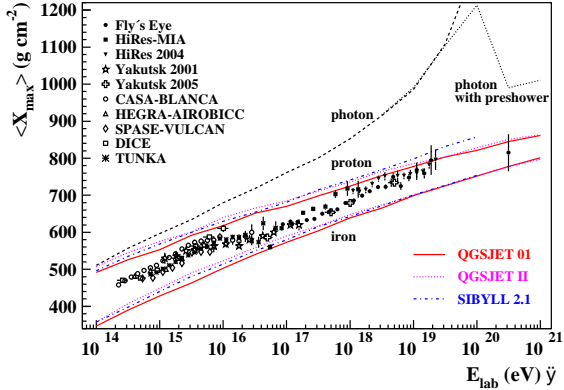}
\caption{Average depth of shower maximum $<$$X_{\rm max}$$>$ versus energy
simulated for primary photons, protons and iron nuclei.
Depending on the specific particle trajectory through the geomagnetic
field, photons above a few times $10^{19}$~eV can also create a
preshower (see e.g.~Ref.~\cite{preshower} and references therein):
as indicated by the splitting of the
photon line, the average $X_{\rm max}$ values then do not only depend
on primary energy but also arrival direction.
For nuclear
primaries, calculations for different hadronic interaction models
are displayed.
One sees that at 10~EeV, photons penetrate deeper into the atmosphere
than hadron primaries by (on average) $\sim$200~g~cm$^{-2}$ or more.
This is to be compared with the resolution of fluorescence telescopes
of typically $\sim$20$-$30~g~cm$^{-2}$.
(Figure taken from Ref.~\cite{augerfd}.)
}
\label{fig1}
\end{figure}

Showers initiated by UHE photons develop differently from showers induced
by nuclear primaries.
Particularly, observables related to the development stage or ``age''
of a shower (such as the depth of shower maximum $X_{\rm max})$
and to the content of shower muons provide good sensitivity to identify
primary photons.
Photon showers are expected to develop deeper in the atmosphere
(larger $X_{\rm max}$), see Fig.~\ref{fig1}.
This is connected to the smaller multiplicity
in electromagnetic interactions compared to hadronic ones, such that
a larger number of interactions is required to degrade the energy
to the critical energy where the cascading process stops.
Additionally, the LPM effect~\cite{lpm} results in a suppression of the
pair production and bremsstrahlung cross-sections.
Photon showers also contain fewer secondary muons, since
photoproduction and direct muon pair production are expected to play
only a sub-dominant role.

\begin{figure}[tb]
\vspace{9pt}
\includegraphics[width=16.5pc]{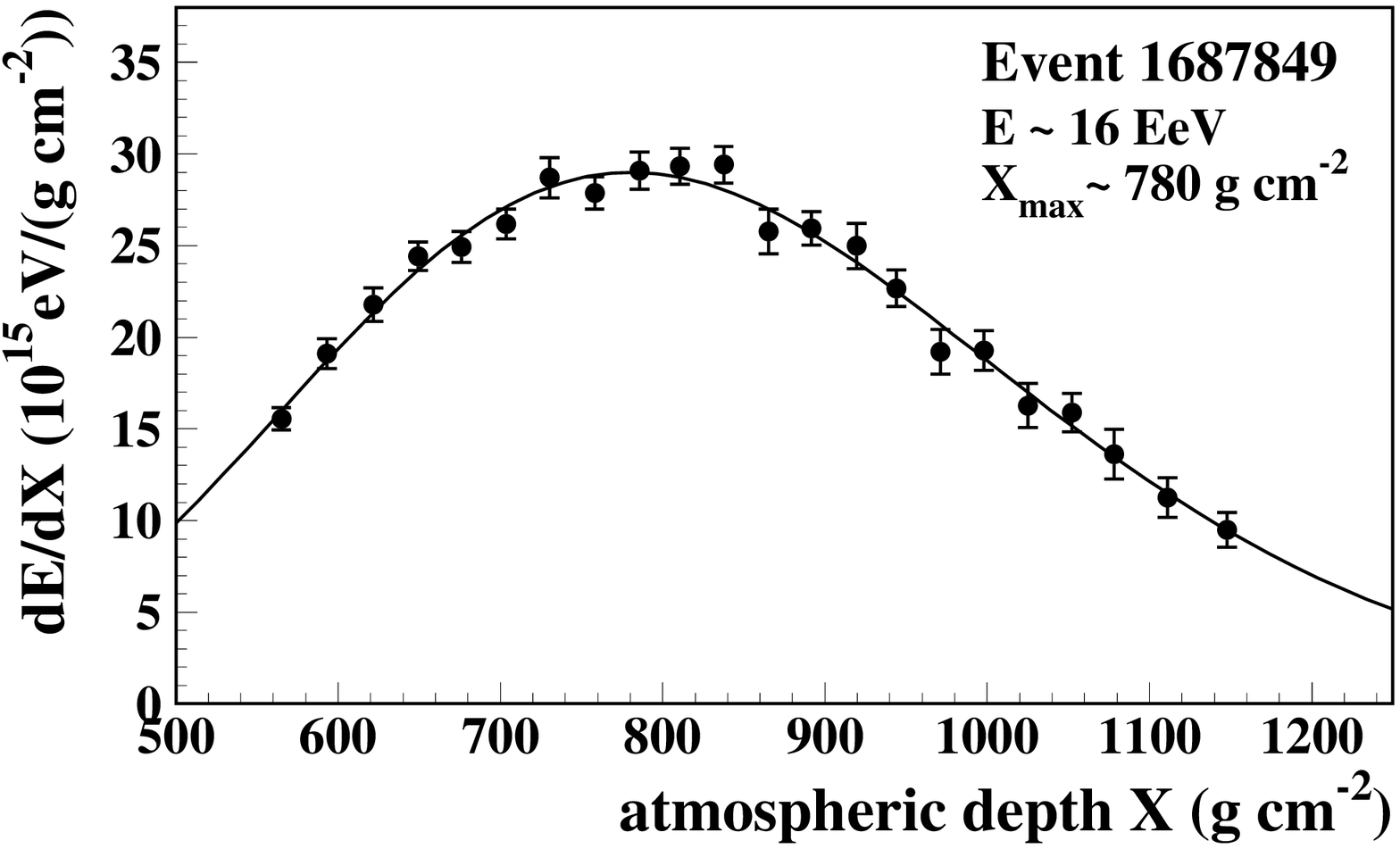}
\includegraphics[width=16.5pc]{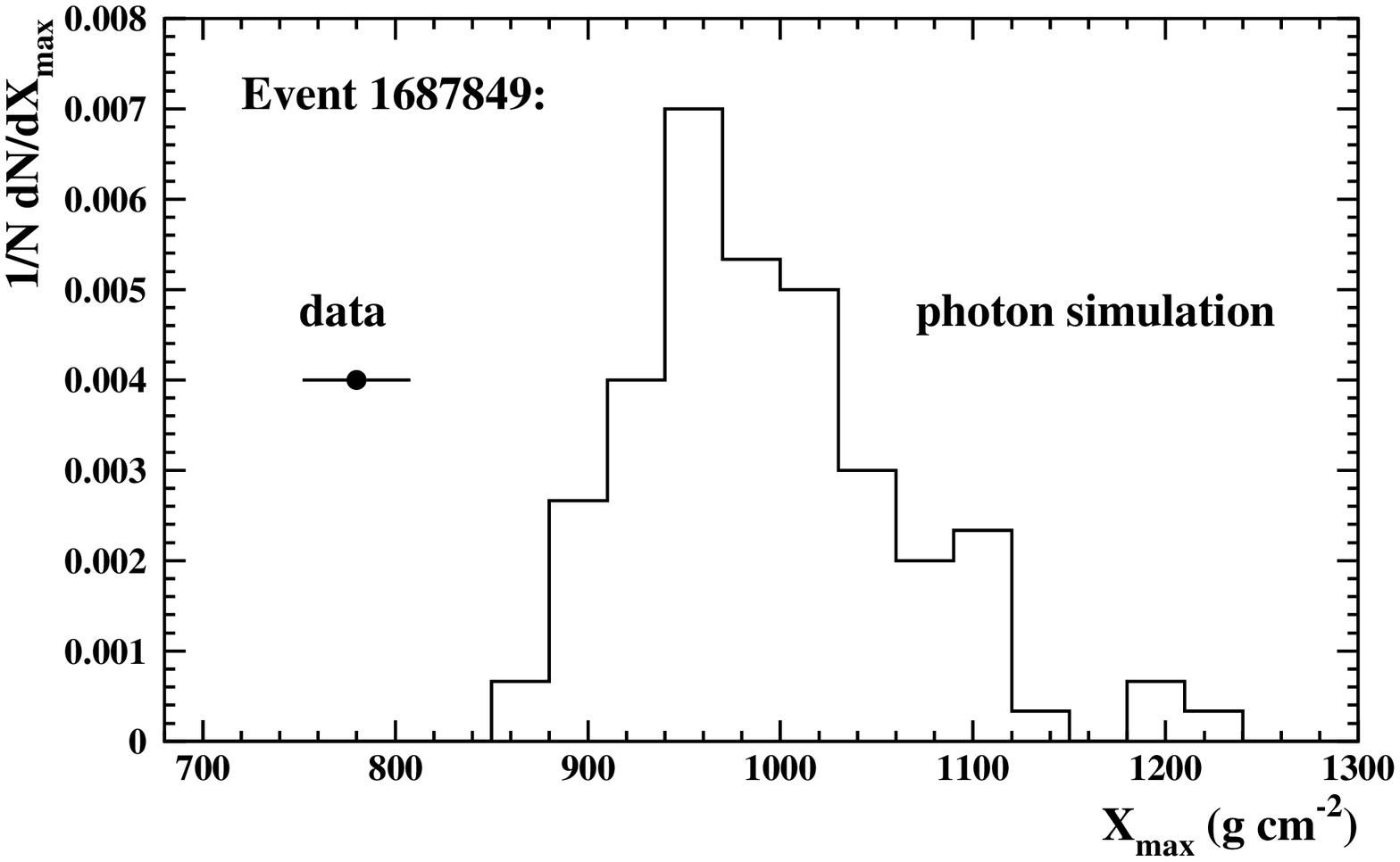}
\caption{
({\it Top}) Example of a reconstructed longitudinal energy deposit profile
(points) and the fit by a Gaisser-Hillas function (line).
({\it Bottom})
$X_{\rm max}$ measured in the shower shown in
the top plot (point with error bar)
compared to the $X_{\rm max}^\gamma$ distribution
expected for photon showers (solid line).
The observed $X_{\rm max}$ is well below the values
assuming primary photons.
(Figures taken from Ref.~\cite{augerfd}.)
}
\label{fig2}
\end{figure}

Using high-quality hybrid events (air showers
registered by both the fluorescence telescopes and the ground array which
fullfil certain strict requirements to be accepted for the analysis),
it is possible to compare directly the observed $X_{\rm max}$ values of UHE events
with expectations for photon primaries of same energy and arrival direction.
An example of a measured air shower and the comparison of its $X_{\rm max}$
value to results from photon simulations is shown in Fig.~\ref{fig2}.
One can see that the observed $X_{\rm max}$ is well below the values
assuming primary photons. It is not likely that this air shower was
initiated by an UHE photon.

Inspecting 29 such high-quality air showers
with energies above 10~EeV, it turned out that in all cases the photon predictions
exceed by far the observed $X_{\rm max}$ values. From this,
an upper limit of 16\% (95\% c.l.) on the
cosmic-ray photon fraction could be obtained as published in 2007~\cite{augerfd}.
Despite the very small
number of events, this limit was the best one at that time above 10~EeV (to be
superseded by the limit using Auger ground array data, see below). Moreover, the Auger
hybrid data set was the first (and so far only) one where the fluorescence
technique with its direct observation of $X_{\rm max}$ was applied to search
for UHE photons.

\section{Search with ground array data}

\begin{figure}[tb]
\vspace{9pt}
\includegraphics[width=19pc]{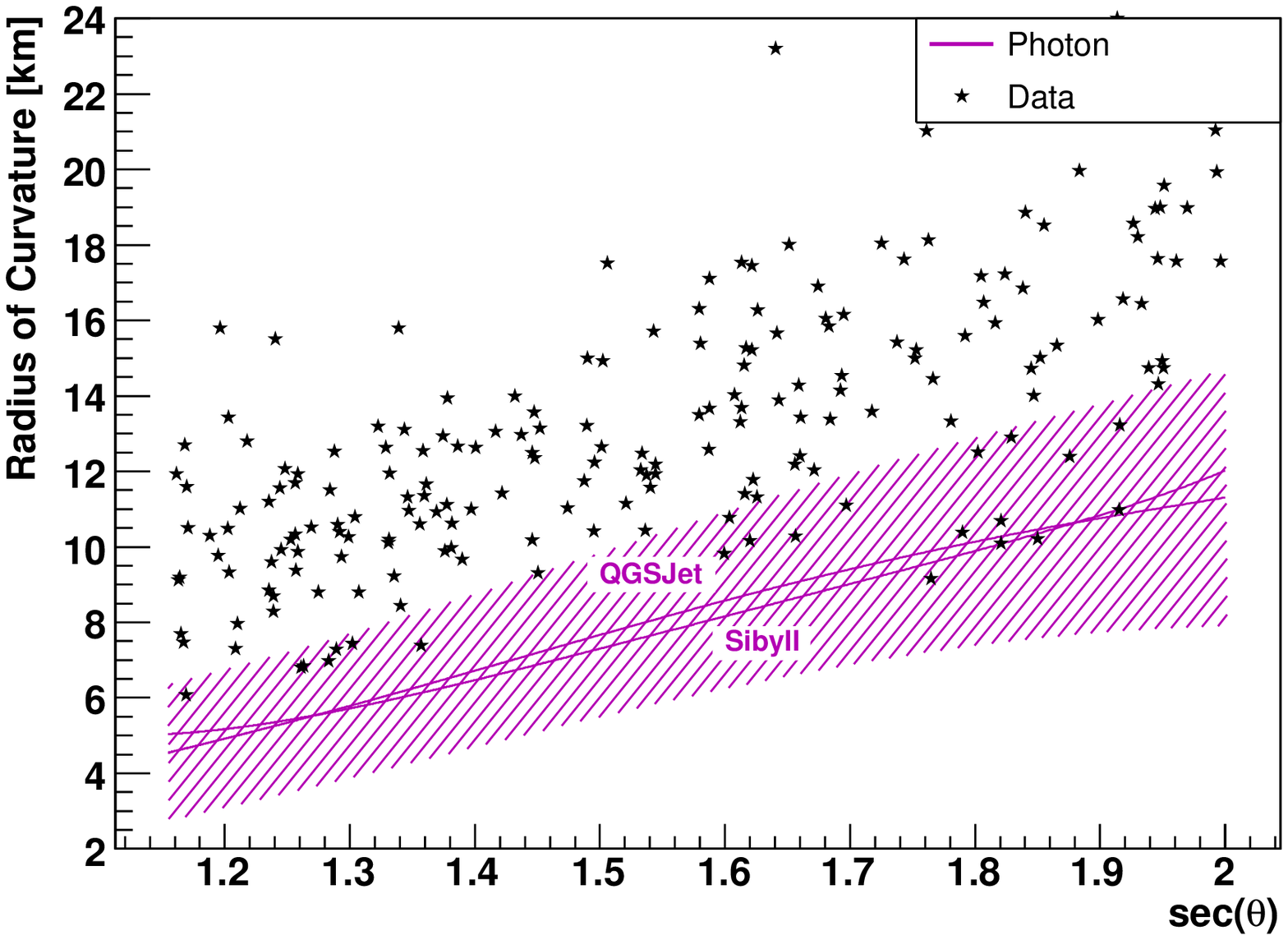}
\includegraphics[width=19pc]{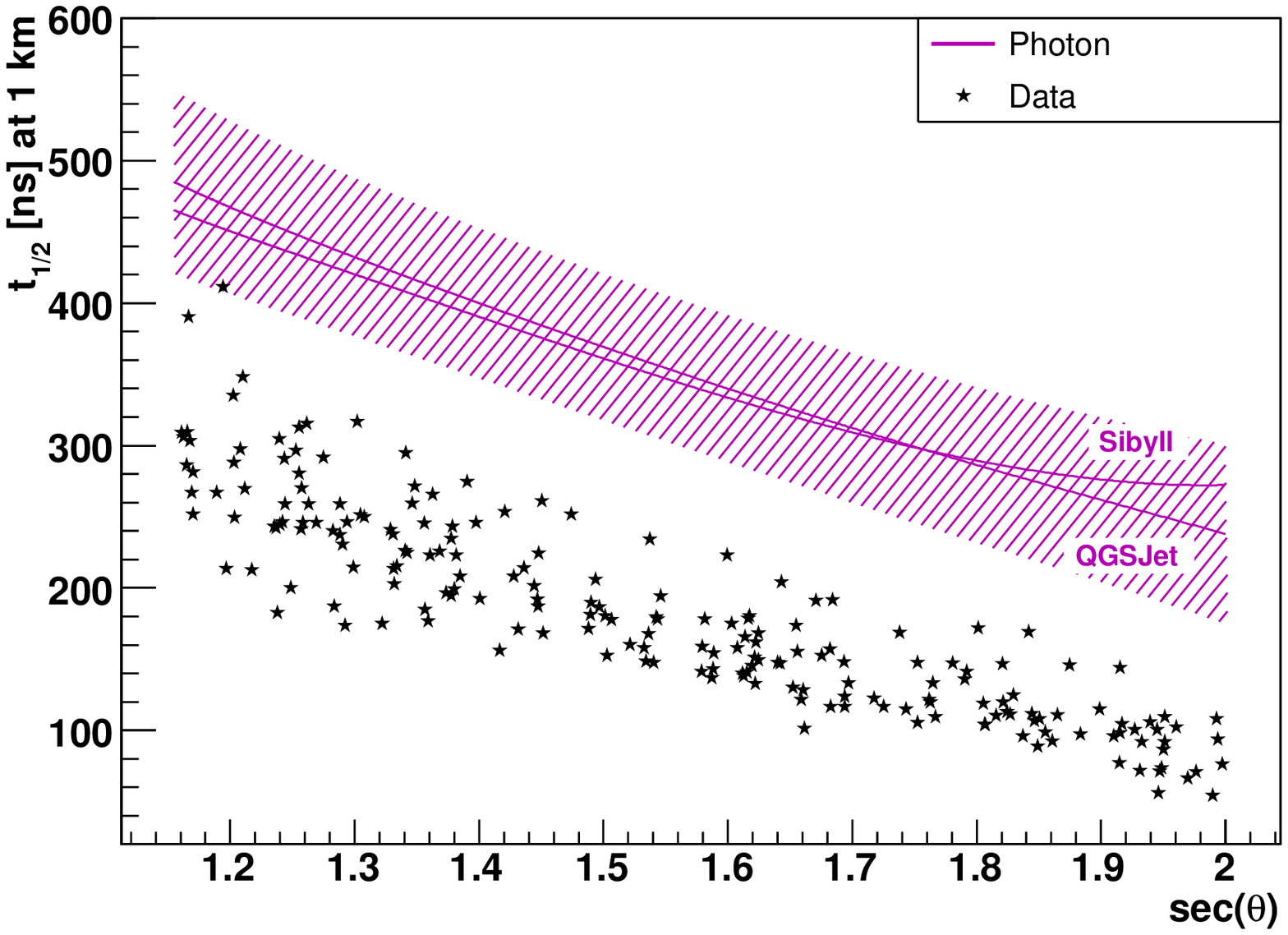}
\caption{Parameterization of the mean behavior of the radius of
  curvature $R$ (upper plot) and shower risetime at 1000~m core distance $t_{1/2}$ 
  (lower plot) for 20~EeV primary photons as a function of the zenith angle
  using two different hadronic interaction models.
  An increase (a decrease) of $R$ (of $t_{1/2}$) with zenith angle
  is expected
  due to the generally longer path lengths to ground in case of
  larger inclination.
  Real events of 19--21~EeV (photon energy scale) are added.
  The significant deviation of the observed values from those expected
  for primary photons is visible.
(Figures taken from Ref.~\cite{augersd}.)
}
\label{fig3}
\end{figure}

\begin{figure}[tb]
\vspace{9pt}
\includegraphics[width=19pc]{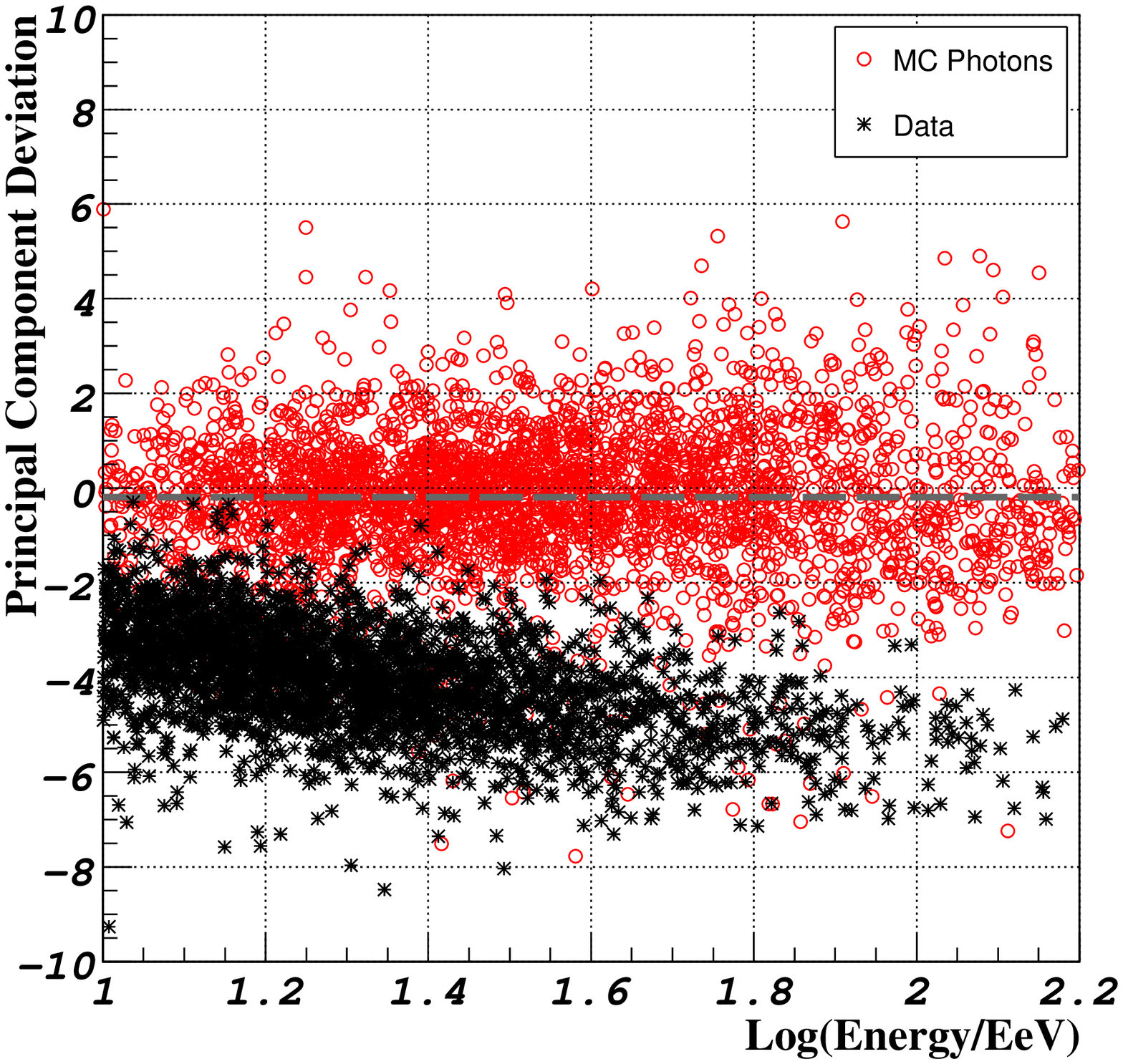}
\caption{Plotted is a quantity that combines the measurements
of radius of curvature and shower risetime (cf.~Fig.~\ref{fig3})
for data (black crosses) and photon simulations (open red circles)
as a function of the primary energy (photon energy scale).
Data lying above the dashed line, which indicates the mean of the
distribution for photons, are taken as photon candidates.
No event meets this requirement.
Moreover, no trend is visible of higher-energy events becoming
more photon-like.
(Figure taken from Ref.~\cite{augersd}.)
}
\label{fig4}
\end{figure}

To exploit the larger number of UHE events recorded by the ground array
only (i.e.~without requiring additional fluorescence telescope data), two
observables of the array detectors were chosen in an analysis published
in 2008~\cite{augersd} which have significantly different
behavior for nuclear primaries when compared to photons:
the risetime of the recorded shower signal and the radius of
curvature of the shower front (see Fig.~\ref{fig3}).
As an energy estimator, the total signal deposit $S(1000)$ derived for
a ground detector at 1000~m core distance is used by comparing the
measured value to photon shower simulations.

In Fig.~\ref{fig4}, for each measured shower a quantity is plotted as a
function of energy of the event that combines the discriminating observables
of risetime and radius of curvature. Also shown are expectations from photon
shower simulations. An {\it a priori} cut had been defined such that data
points lying above the mean of the photon distribution are considered as
photon candidates. Accounting for inefficiencies, upper limits on the
presence of photons in the primary cosmic-ray beam were placed; in
particular a limit of 2\% (at 95\% c.l.) above 10~EeV was derived
(see Fig.~\ref{fig5}, lower plot).

Also, in this analysis the first direct limit to the {\it flux} of UHE photons
(instead of the {\it fraction}) was obtained (Fig.~\ref{fig5}, upper plot).
In case of ground arrays, the
flux (or a limit on the flux) is the more robust experimental quantity
(see Refs.~\cite{augersd,review}).

\section{Implications}

\begin{figure}[tb]
\vspace{9pt}
\includegraphics[width=19pc]{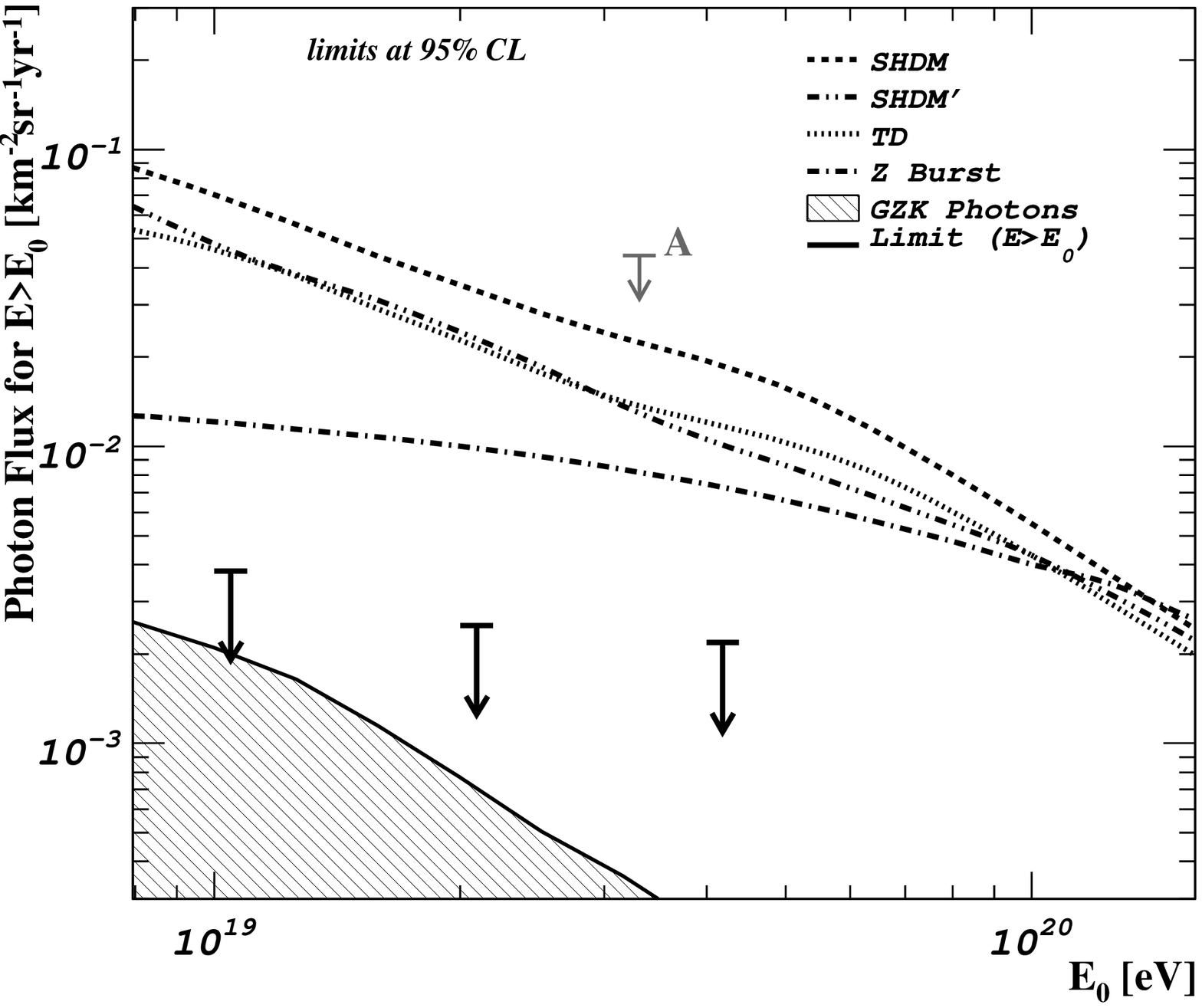}
\includegraphics[width=19pc]{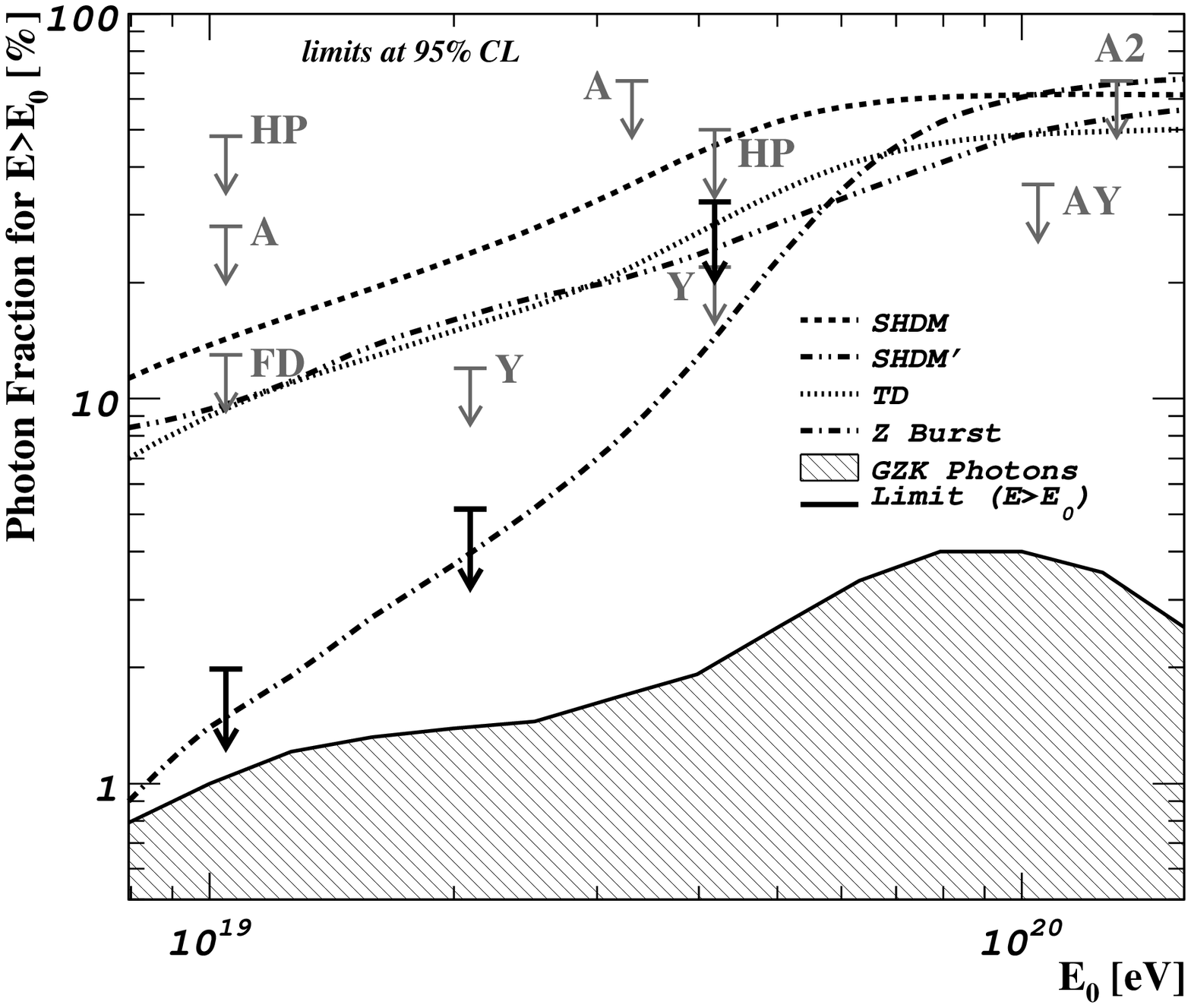}
\caption{The upper limits on the integral flux of photons
{\it (top)} and on the fraction of photons
  in the integral cosmic-ray flux {\it (bottom)} as a
function of the threshold energy as
measured by the Auger array detector (black arrows)
along with previous experimental limits and
model predictions.
(Figures taken from Ref.~\cite{augersd}.)
}
\label{fig5}
\end{figure}

As can be seen from Fig.~\ref{fig5}, particularly from the comparison
of experimental UHE photon {\it flux} limits and model predictions, contemporary
top-down models
are severely constrained now: the non-observation of UHE photons by the
Auger Observatory is difficult to explain in exotic physics scenarios
(see e.g.~also Refs.~\cite{aloisio,kachelriess}).

The photon limits also reduce systematic uncertainties in the derivation
of the total cosmic-ray flux spectrum~\cite{spectrum}, since the energies
of photon showers could be misreconstructed. As can also be seen
in Fig.~\ref{fig4}, there are no indications of an increasing photon
component towards highest energies at the present sensitivity level.

The photon bounds have also proven useful for fundamental physics.
In Ref.~\cite{liv}, the fraction of {\it GZK photons} expected at Earth
was computed assuming that Lorentz invariance violation suppresses
pair production by UHE photons. Values were found that exceed the Auger
photon limits by about a factor 10.
Corresponding constraints on Lorentz violation
parameters improve previous constraints by several orders of magnitude
due to the extreme energy in case of UHE photons.

\section{Outlook}

Data taken during the first phase of construction of the southern Auger
Observatory allowed the derivation of UHE photons limits that approach
(in terms of fraction) the $10^{-2}$ level at energies above 10 EeV. 
Both detector components (telescopes and array) offer excellent
discrimination power between photon and hadron primaries such that the
sensitivity level will continue to improve over the next years while the
data accumulate. The upper range of predictions for GZK photons (see
Fig.~\ref{fig5}) as well as the yet unexplored energy range below
10~EeV can be tested soon. 

With a main goal of full sky coverage, the Auger Observatory is to be
completed by a northern site. Current plans aim at a significantly
(factor $\sim$7) larger array to proceed with cosmic-ray astronomy.
With such an enlarged northern site, photon fractions at or below the
$10^{-3}$ level are in reach within few years of operation.

As for any exploratory search, the timescale for the final discovery
of UHE photons is uncertain.
However, even the very first results from the photon search at the
Auger Observatory proved very helpful for cosmic-ray physics
(e.g.~discrimination between different cosmic-ray source scenarios)
and for fundamental physics (test of Lorentz invariance).
It seems reasonable to believe that this ability of the UHE photon
search to provide substantial physics results will persist.
The unprecedented and presently {\it unique} sensitivity of
the Auger Observatory to UHE photons, particularly when completed by
the northern site, allows for the first time a realistic search for
UHE photons also from conventional cosmic-ray scenarios. 

A final discovery of UHE photons would open a new $-$ and
the most extreme $-$ ``window'' of photon astronomy.
Experience shows that this is usually accompanied
with radically new, and often unexpected, insights
about the bizarre inhabitants of the universe and their 
(inter-)actions~\cite{lawrence} (for a surely incomplete list of
possible impacts of a discovery of UHE photons on various research
fields, see Ref.~\cite{review}).
\\

{\it Acknowledgments.}
It is a pleasure to thank the {\it CRIS} organizers for an
excellent and inspiring conference. (The sometimes quite ``German''
weather made me feel at home even more.)
Partial support from the German Ministry for Education and
Research BMBF (Verbundforschung Astro\-teilchenphysik) and
from the DFG are kindly acknowledged.


\begin{thebibliography}{9}

\bibitem{auger}
J.~Abraham et al. (Pierre Auger Collaboration),
Nucl.~Instrum.~Meth.~{\bf A 523}, 50 (2004).

\bibitem{topdown}
P.~Bhattacharjee, G.~Sigl, Phys.~Rep.~{\bf 327}, 109 (2000).

\bibitem{gzk-photons}
G.~Gelmini, O.~Kalashev, D.V.~Semikoz, [arXiv:astro-ph/0706.2181];
G.~Sigl, Phys.~Rev.~D {\bf 75}, 103001 (2007) [arXiv:astro-ph/0703403].

\bibitem{augerfd} J.~Abraham et al. (Pierre Auger Collaboration),
   Astropart. Phys. {\bf 27} (2007) 155

\bibitem{augersd} J.~Abraham et al. (Pierre Auger Collaboration),
   Astropart. Phys. {\bf 29} (2008) 243

\bibitem{review} M. Risse, P. Homola, Mod. Phys. Lett. A {\bf 22}, 749 (2007).

\bibitem{lpm}
L.D.~Landau, I.Ya.~Pomeranchuk,
Dokl. Akad. Nauk SSSR {\bf 92}, 535 \& 735 (1953);
A.B.~Migdal, Phys. Rev. {\bf 103}, 1811 (1956).

\bibitem{preshower} P.~Homola et al., Comp.~Phys.~Comm.~{\bf 173}, 71 (2005)

\bibitem{aloisio} R. Aloisio, F. Tortorici, Astropart. Phys. {\bf 29}, 307 (2008).

\bibitem{kachelriess} M. Kachelrie{\ss}, arXiv:0810.3017v2

\bibitem{spectrum} Pierre Auger Collaboration, Phys.~Rev.~Lett. {\bf 101}, 061101 (2008)

\bibitem{liv} M. Galaverni, G. Sigl, Phys. Rev. Lett. {\bf 100}, 021102 (2008).

\bibitem{lawrence} A. Lawrence, arXiv:0704.0809



\end{thebibliography}
\end{document}